\documentclass{webofc}
\usepackage[varg]{txfonts}   
\begin{document}
\title{Emergence of ghost in once-subtracted on-shell unitarization in glueball-glueball scattering}
%
%

\author{Enrico Trotti\inst{1}\fnsep\thanks{\email{trottienrico@gmail.com}} 
}

\institute{Institute of Physics, Jan Kochanowski University,
ul. Uniwersytecka 7, 25-406, Kielce, Poland 
          }

\abstract{ We investigate the scattering of two scalar glueballs in pure YM theory, using the well known dilaton potential. We perform the calculations considering a glueball mass of about $m_G \approx 1.7$ GeV, as predicted by lattice QCD. We begin with the tree-level theory, but the question about the presence of a bound state needs a deeper study to be answered. Thus we unitarize the theory through a self energy loop function consisting of a single subtraction at the single glueball resonance pole. We show that this choice is inconsistent as it leads to the emergence of a ghost-like state with negative norm. This problem is related with the sign of the coefficient in the first order term of the expansion of the reverse unitarized amplitude. We briefly discuss the solution which consists of an additional subtraction in the loop function, as presented in Eur.Phys.J.C 82 (2022) 5, 487.
}
\maketitle
\section{Introduction}
\label{intro}

The theory of QCD in the limit of infinitely heavy quarks -Yang-Mills (YM) theory- provides several predictions for the glueballs using different approaches, first of all the bag models \cite{Chodos:1974je,Jaffe:1975fd} and later Lattice QCD \cite{Chen:2005mg,Morningstar:1999rf,Caselle:2001im,Athenodorou:2020ani}.
Gluons are the gauge bosons of QCD and their bound states are glueballs. It is therefore natural to investigate the bound state of glueballs, in their turn. The lightest glueball state in all approaches mentioned above is the scalar one, $J^{PC}=0^{++}$, making it stable in pure YM. \\ 
In our work, we consider the dilaton potential \cite{Migdal:1982jp,Salomone:1980sp}, which involves a single scalar glueball field $G$ and a single dimentionful parameter $\Lambda_G$ whose value, obtained by comparison with the gluon condensate as determined on the lattice, is about $0.4$ GeV. 
The scattering amplitudes obtained from the potential are necessary to access if the attraction between the two particles is enough to form a bound state. 
The tree-level theory does not provide satisfying results and must be therefore unitarized via a loop function. The first attempt with a single subtracted loop provides a ghost like state as a result, which is clearly an artifact of the theory.
The unitarized theory with a double subtraction, instead, gives an interesting result: a bound state, that we call glueballonium, might exist below a critical value of the parameter $\Lambda_G$ -$\Lambda_{G,crit} \approx 0.504$ GeV-, as it was more exhaustively explained in Refs. \cite{Trotti1, Trotti2}. \\
In the future, the results of our work can be compared with lattice QCD, since the scattering of two glueballs in pure YM can be studied using the Lusher method \cite{Andersen:2018mau,Luscher:1990ux}. In full QCD, the situation is more difficult since the scalar glueball could be broad \cite{Ellis:1984jv}, yet a search for the glueballonium at the PANDA experiment would be possible \cite{PANDA:2009yku}. Moreover, a deeper understanding of the dilaton potential on its own can be also relevant beyond QCD \cite{Goldberger:2007zk}, e.g. dark matter \cite{Yamanaka:2021xqh,Yamanaka:2019aeq,Yamanaka:2019yek,Yamanaka:2019gak,Tseytlin:1991ss,Gasperini:1994kb}.

\section{Scattering at tree level}
\label{TL}

A theory which involves a single scalar field $G$ and one dimentional parameter $\Lambda_G$, embodying the trace anomaly at the hadronic level, is the main input in this work.
The dilaton Lagrangian well describes the low-energy features of the YM sector of QCD \cite{Migdal:1982jp}:
\begin{equation}%
\mathcal{L}%
_\text{dil}=\frac{1}{2}(\partial_{\mu}G)^{2}-V(G),
\label{lagra}
\end{equation}
with%
\begin{equation}
V(G)=\frac{1}{4}\frac{m_{G}^{2}}{\Lambda_{G}^{2}}\left(  G^{4}\ln\left\vert
\frac{G}{\Lambda_{G}}\right\vert -\frac{G^{4}}{4}\right)  \text{.}%
\label{potential}%
\end{equation}
The term $m_G$, which appears in the dimensionless quantity $m_G/\Lambda_G$, represents the glueball mass, which is fixed to $m_G \approx 1.7$ GeV in this paper \cite{Morningstar:1999rf,Athenodorou:2020ani,Sexton:1995kd,Giacosa:2005zt,Gui:2012gx,Chen:2005mg,Janowski:2014ppa}. The expansion of the potential $V(G)$ around its minimum, realized for $G=\Lambda_G$,  gives the expression of the 2$\rightarrow$2-body scattering amplitude at tree level, formed by the cross-channel and the $s,t,u$ channels:
\begin{equation}
A(s,t,u)=-11\frac{m_{G}^{2}}{\Lambda_{G}^{2}}-\left(  5\frac{m_{G}^{2}%
}{\Lambda_{G}}\right)  ^{2}\frac{1}{s-m_{G}^{2}} 
-\left(  5\frac{m_{G}^{2} 
}{\Lambda_{G}}\right)  ^{2}\frac{1}{t-m_{G}^{2}}   -\left(  5\frac{m_{G}^{2}%
}{\Lambda_{G}}\right)  ^{2}\frac{1}{u-m_{G}^{2}} \text{ ,}\label{totampl}%
\end{equation}
which is expressed as function of the Mandelstam variables $s = (p_{1} + p_{2})^{2}$, $t=(p_{1}-p_{3})^{2}$ and $u=(p_{2}-p_{3})^{2}$. The amplitude can be rewritten as function of $s$ and the scattering angle $\theta$, as only two of the Mandelstam variables are independent. The expansion into partial waves is expressed via a general formula, that for the $l$-th amplitude reads:
\begin{equation}
    A_{l}(s)=\frac{1}{2}\int_{-1}^{1}d\cos\theta A(s,\cos\theta)P_{l}(\cos\theta) \text{ .}\label{l-ampl}
\end{equation}
Because of Bose symmetry, $A(s,\cos\theta)$ is symmetric in $\cos\theta$; therefore, odd waves vanish. Additionally, waves with $l\geq 2$ do not contribute to the formation of the bound state, thus the most relevant value of $l$ to consider is $l=0$. 
The explicit form of the s-wave amplitude is:

\begin{equation}
A_{0}(s)
=-11\frac{m_{G}^{2}}%
{\Lambda_{G}^{2}}-25\frac{m_{G}^{4}}{\Lambda_{G}^{2}}\frac{1}{s-m_{G}^{2}%
}+50\frac{m_{G}^{4}}{\Lambda_{G}^{2}}\frac{\log\left(  1+\frac{s-4m_{G}^{2}%
}{m_{G}^{2}}\right)  }{s-4m_{G}^{2}}\,. \label{eq:A0}%
\end{equation}
The form of the amplitude directly shapes other quantities e.g. scattering length $a_l$ and phase shift $\delta_l$, which expression and discussion are in Ref. \cite{Trotti2,Samanta:2020pez}.
The pole at $s=m_G^2$ in Eq. \ref{eq:A0} can be immediately interpreted as the single glueball. Interestingly, the amplitude $A_0(s)$ presents an additional singularity at $s=3m_G^2$, which can be understood by comparison between Eq. (\ref{totampl}) and the last term of Eq. (\ref{eq:A0}): the projection of the $u$- and $t$- channel single pole onto the s-wave generates a left hand cut, embodied in the logarithmic term. 

\section{On shell unitarization}
\label{unitar}

\subsection{General unitarization features}
\label{unitargeneral}

As the tree-level amplitude works quite well only within a small neighbourhood of the threshold and for weak couplings, it is necessary to consider also loops. To this end, we will use a procedure called unitarization.
There is not a unique way for unitarizing any $l$-th wave amplitude obtained from a chiral lagrangian \cite{Truong:1991gv,GomezNicola:2001as,Cudell:2008yb}. Two ways are via the "on-shell approximation" \cite{Oller:1997ng} or the N/D method \cite{Gulmez:2016scm,Cahn:1983vi}. A discussion of the latter can be found in Ref. \cite{Trotti2}; from now on the former will be the focus of our discussion. The on-shell approximation is realized with the help of a glueball-glueball self energy loop function (or vacuum polarization function) $\Sigma(s)$, as follows:
\begin{equation}
A_{0}^{\text{U}}(s)=\left[  A_{0}^{-1}(s)-\Sigma(s)\right]  ^{-1}= \dfrac{A_{0}(s)}{1-A_{0}(s)\Sigma(s)} ,
\label{eqaluni}%
\end{equation}
where the tree-level $A_0(s)$ is directly related with the unitarized amplitude $A_{0}^{\text{U}}(s)$. It is evident from the denominator that the loop contribution in case of small value of $A_0(s)$ is negligible. Thanks to the optical theorem, the imaginary part of $\Sigma(s)$ must have the following form \cite{Giacosa:2007bn}:
\begin{equation}
\text{Im}\Sigma(s)=\theta\!\left(s - 4m_G^2\right)
\frac{1}{2}\frac{1}{16\pi}\sqrt{1-\frac{4m_{G}^{2}}{s}}\label{imloop} \text{ .}%
\end{equation}
The imaginary part can be directly used to reconstruct the whole loop function by using the dispersion relations, which can require a certain number of subtraction in order to preserve some analytic features during the unitarization process. In this way, $\Sigma(s)$ takes the general form:
\begin{equation}
\Sigma(s)= \dfrac{\prod_{j}(s-\mu_j^2)}{\pi}\int_{4m_G^{2}}^{\infty}%
\frac{\text{Im} \Sigma(s^{\prime})}{(s^{\prime}-s-i\varepsilon)\prod_j (s^{\prime}-\mu_j^2)}ds^{\prime}\text{ ,} \label{loop}%
\end{equation}
where $\mu_j^2$ are the positions of the poles that we want to preserve during the unitarization, so that $\Sigma(s=\mu_j^2)=0$ for each $\mu_j^2$.
The first pole that we keep fixed is the single glueball pole $\mu_j^2=m_G^2$, as unitarization must preserve the physical mass (assumed to be such already at tree-level). One pole is sufficient to ensure the convergence of the loop function, but the formalism leads to an unphysical ghost pole, see the following. The solution to this problem lies in the addition of another pole. The requirement to preserve the single glueball exchange also in the $t$- and $u$- channel, i.e. $A_{0}^{\text{U}}(s)\simeq A_{0}(s)$ near the logarithmic branch point, leads to a new subtraction at $\mu_j^2=3m_G^2$. \\
Next, we will analyse which are the features and the problems of the theory in case of a single subtraction, as the double subtracted case has been exhaustively explained in Refs. \cite{Trotti2,Trotti1}. In particular, we show the emergence of a ghost particle.

\subsection{Once subtracted loop function}
\label{subonce}

\begin{figure}[h]
\centering
\includegraphics[width=12cm,clip]{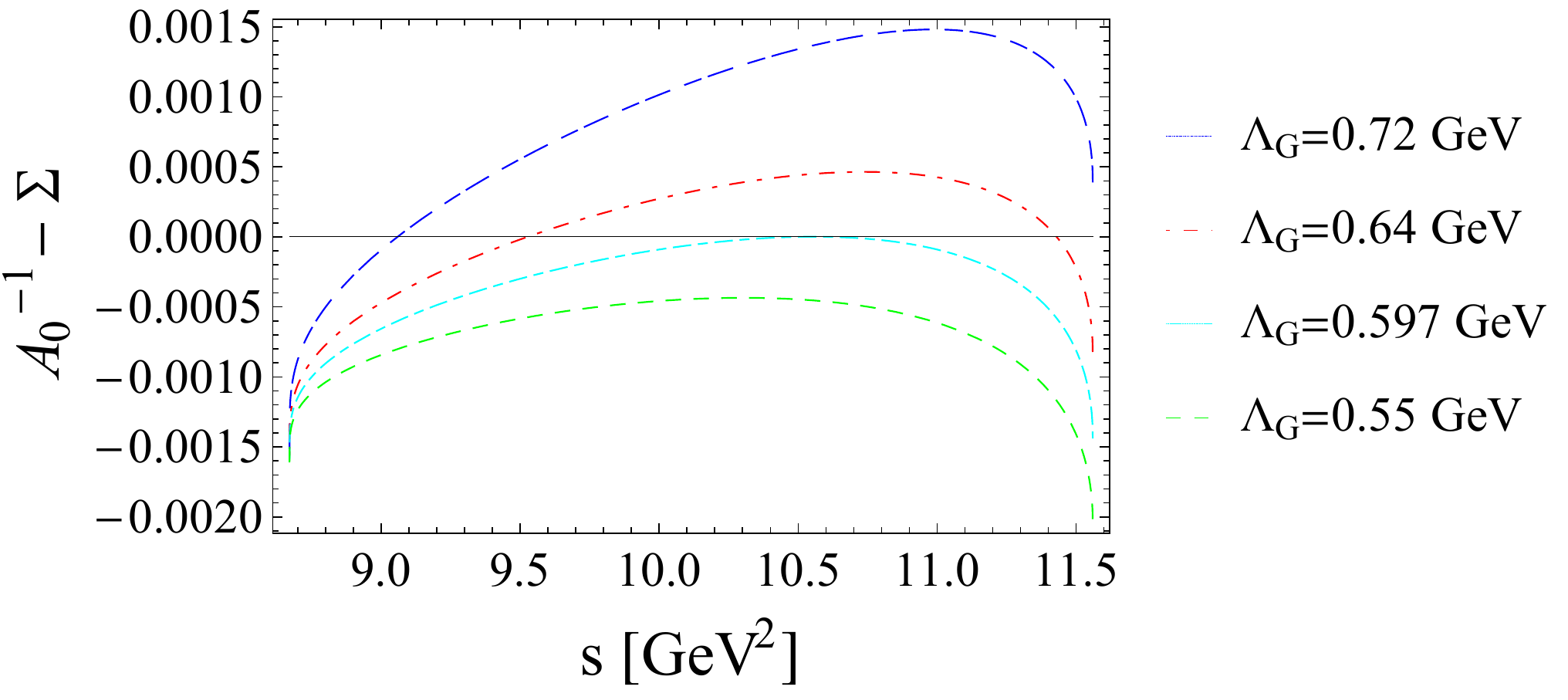}
\caption{The denominator of the unitarized amplitude $(A_0^U)^{-1}=A_0^{-1}- \Sigma$ as function of the energy for different values of the scale $\Lambda_G$.}
\label{fig-1}       
\end{figure}

We consider the loop function with one subtraction in order to preserve the single glueball pole. Thus Eq. \ref{loop} becomes
\begin{equation}
\Sigma(s)= \dfrac{(s-m_G^2)}{\pi}\int_{4m_G^{2}}^{\infty}%
\frac{\text{Im} \Sigma(s^{\prime})}{(s^{\prime}-s-i\varepsilon)(s^{\prime}-m_G^2)}ds^{\prime}\text{ .} \label{looponce}%
\end{equation}
The presence of a bound state is realized as a pole of the amplitude. It is therefore evident from Eq. \ref{eqaluni} that we have to analyse for which values of $s$:
\begin{equation}
    A_{0}^{-1}(s,\Lambda_G)=\Sigma(s).
    \label{equivalence}
\end{equation}
Note, we consider the scale $\Lambda_G$ as a variable; indeed, as shown in Fig. \ref{fig-1}, the unitarized amplitude has two, one or no pole depending on the value of the scale. In the case of a too strong attraction (that is, relatively small $\Lambda_G$) -e.g. $\Lambda_G=0.55$ GeV-, there is no pole.  Weakening the coupling, we firstly see that a pole appears at $s \approx 3.65$ GeV$^2$, and then it splits into two singularities, one for a lower and another for a higher value of $s$. The latter approaches $s=4m_G^2$ up to $\Lambda_G \approx 0.69$ GeV, and then disappear. This behaviour is proper of a bound state singularity.
Instead, the pole at lower $s$ does not disappear when the strength of the coupling decreases (alias, $\Lambda_G$ increases) and it is present even for the quite unrealistic value of $\Lambda_G=1.5$ GeV. In order to solve this problem, it is therefore necessary to understand the nature of this pole, which will be done in an analogous way as reported in Ref. \cite{Donoghue:2019fcb}. 
We expand the denominator of Eq. \ref{eqaluni} near a certain pole  $s \approx \mu_j^2$:
\begin{equation}
    A^{-1}-\Sigma \approx O [(s-\mu_j^2)^0] + C(s-\mu_j^2)+ O[(s-\mu_j^2)^{n \geq 2}].
\end{equation}
The coefficient $C$ is related to the sign of the first derivative of the term $A^{-1}-\Sigma$; the sign of $C$ fixes the properties of the coupling constant. The case $C \textless 0$ corresponds to a bound state -$C=1/g^2_{eff,b}$-, while $C \textgreater 0$ embodies a ghost state -$C=1/g^2_{eff,g}$-; the indexes $b$ and $g$ refer to bound state and ghost respectively.
The unitarized amplitude gets then two forms, depending if it is near the bound state pole, $\mu_j=m_b$, or the ghost pole, $\mu_j=m_g$:
\begin{equation}
A_{0}^{\text{U}}(s\approx m_b^2)=-\dfrac{g_{eff,b}^2}{s-m_b^2} 
\label{eq12}
\end{equation}
and
\begin{equation}
A_{0}^{\text{U}}(s\approx m_g^2)=\dfrac{g_{eff,g}^2}{s-m_g^2}.
\label{eq13}
\end{equation}
The two amplitudes can be obtained by the corresponding  effective Lagrangians for two glueball fields into the effective fields $\phi_b$ and $\phi_g$:
\begin{equation}%
\mathcal{L}_{eff,b}=\dfrac{1}{2}(\partial_{\mu}\phi_b)^2-\dfrac{m_b^2}{2}\phi_b^2+  g_{eff,b}G^2\phi_b.
\label{efflagrab}
\end{equation}
and
\begin{equation}%
\mathcal{L}_{eff,g}=\dfrac{1}{2}(\partial_{\mu}\phi_g)^2-\dfrac{m_g^2}{2}\phi_g^2+ i g_{eff,g}G^2\phi_g.
\label{efflagrag}
\end{equation}
Clearly, $\mathcal{L}_{eff,g}$ is not Hermitian. In order to have an Hermitian $\mathcal{L}_{eff,g}$, one could consider the transformation $\phi_g \rightarrow i\phi_g$.
The Lagrangian in Eq. \ref{efflagrag} gets the form:
\begin{equation}%
\overline{\mathcal{L}}_{eff,g}=-\dfrac{1}{2}(\partial_{\mu}\phi_g)^2+\dfrac{m_g^2}{2}\phi_g^2- g_{eff,g}G^2\phi_g.
\label{efflagrag2}
\end{equation}
Although $\overline{\mathcal{L}}_{eff,g}$ is now Hermitian, the sign in front of the kinetic part is reversed. Thus, any formulation of the theory leads to a state with negative norm for $s \approx m_g^2$.

\noindent
In the case of this work, we can consider the poles of the unitarized amplitude at $\Lambda = 0.64$ GeV (see Fig.\ref{fig-1}). The coefficient $C$ is then about $7.1 \cdot 10^{-4}$ GeV$^{-2}$ for $m_g \approx 3.08$ GeV, meaning that this is an unphysical ghost-like resonance. The other pole at $m_b \approx 3.38$ GeV has a negative value of $C$ ($\approx - 2 \cdot 10^{-3}$ GeV$^{-2}$), meaning that it is the bound state of the theory. In conclusion, one subtraction is not enough. \\
The theory must be modified with the addition of at least another subtraction to the loop function, see details in Ref. \cite{Trotti2}.

\section{Conclusions}
In this work, we used the dilaton potential to study the scattering of two scalar glueballs. Since the tree-level amplitude is not a sufficient approximation, a unitarization process was implemented, which allows us to take into account loop contributions. We implemented the calculation with the on-shell formalism, which employs the self energy function; its real part is only fixed after a certain number of subtractions is implemented. \\
The case with one only subtraction at the single glueball pole was analysed in this work. We have shown that this choice does not provide a satisfactory unitarization for the scattering of two scalar glueballs, as it generates, together with the bound state pole, an unphysical ghost-like resonance. The problem can be solved by considering an additional subtraction at the branch cut.

\section*{Acknowledgments}
The author thank F. Giacosa and A. Pilloni with whom Ref. \cite{Trotti2}  was written. 
The author acknowledges financial support through the project AKCELERATOR ROZWOJU
Uniwersytetu Jana Kochanowskiego w Kielcach (Development Accelerator of the
Jan Kochanowski University of Kielce), co-financed by the European Union
under the European Social Fund, with no. POWR.03.05.00-00-Z212/18.

\bibliographystyle{apsrev4-2.bst}
\bibliography{quattro}



%

%

\end{document}